\documentclass{article}

\usepackage{arxiv}

\usepackage[utf8]{inputenc} % allow utf-8 input
\usepackage[T1]{fontenc}    % use 8-bit T1 fonts
\usepackage{hyperref}       % hyperlinks
\usepackage{url}            % simple URL typesetting
\usepackage{booktabs}       % professional-quality tables
\usepackage{amsfonts}       % blackboard math symbols
\usepackage{nicefrac}       % compact symbols for 1/2, etc.
\usepackage{microtype}      % microtypography
\usepackage{graphicx}
\usepackage{array}
\usepackage{tabularx}
\usepackage{ragged2e}
\usepackage{natbib}
\usepackage{doi}

\title{AI-PACE: A Framework for Integrating AI into Medical Education}

\author{
  \href{https://orcid.org/0000-0002-0065-726X}{\includegraphics[scale=0.06]{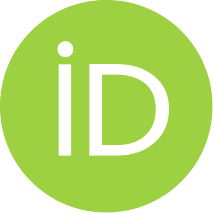}\hspace{1mm}Scott P.~McGrath}\thanks{Corresponding author: \texttt{smcgrath@berkeley.edu}} \\
  Center for Information Technology in the Interest of Society \\
  University of California, Berkeley \\
  Berkeley, CA, USA \\
  \texttt{smcgrath@berkeley.edu} \\
  \And
  Katherine K.~Kim \\
  School of Medicine, Department of Public Health Sciences \\
  University of California, Davis \\
  Davis, CA, USA \\
  \And
  Karnjit Johl \\
  School of Medicine, Department of Internal Medicine \\
  University of California, Davis \\
  Davis, CA, USA \\
  \And
  Haibo Wang \\
  Research Centre of Big Data and AI for Medicine \\
  First Affiliated Hospital of Sun Yat-Sen University \\
  Guangzhou, China \\
  \And
  Nick Anderson \\
  School of Medicine, Department of Public Health Sciences \\
  University of California, Davis \\
  Davis, CA, USA \\
}

\renewcommand{\shorttitle}{AI-PACE: Integrating AI into Medical Education}

\hypersetup{
  pdftitle={AI-PACE: A Framework for Integrating AI into Medical Education},
  pdfauthor={Scott P. McGrath, Katherine K. Kim, Karnjit Johl, Haibo Wang, Nick Anderson},
  pdfkeywords={Medical Education, Artificial Intelligence, Curriculum Development, Implementation Strategy, Bloom's Taxonomy},
}

\begin{document}
\maketitle

\begin{abstract}
Medical AI education remains fragmented, specialty-skewed, and lacks longitudinal structure, particularly for generalist physicians. Through an integrative review of 23 peer-reviewed articles (2016--2025), we identified three structural gaps: short-term interventions without reinforcement, procedural-field bias, and consistent under-representation of the Affective domain. We present AI-PACE (Psychomotor, Affective, Cognitive, Embedded), a Bloom's Taxonomy-grounded framework organizing AI competencies longitudinally across undergraduate, graduate, and continuing medical education.
\end{abstract}


\section{Introduction}
\label{sec:introduction}

Artificial intelligence's (AI) role in medicine is rapidly changing; it is now an active clinical participant, from diagnostic imaging to administrative workflow optimization, compared to an occasional aide in the past. This educational urgency is amplified by how rapidly consumers have adopted generative AI (GenAI). When released in November of 2022, ChatGPT set the record for the fastest-growing user base, eclipsing previous record holders like TikTok and Instagram~\citep{hu2023chatgpt}. With GenAI, patients are not waiting for clinicians, they are forging ahead of healthcare adoption~\citep{salmi2026healthgpt, stokelwalker2024patients}. They are already querying current models from OpenAI, Google, and Anthropic about medical conditions.

OpenAI stated that in January of 2026, 230 million customers were asking ChatGPT about health and wellness per week~\citep{silberling2026chatgpthealth}. Both OpenAI and Anthropic have now released health-focused products in reaction to these market demands (Claude for Healthcare~\citep{anthropic2026healthcare} and ChatGPT Health~\citep{openai2026chatgpthealth}). It is now possible for physicians to encounter patients who could activate AI voice assistants mid-visit to interpret recommendations in real-time. Hence, we are facing a ``shadow'' health system that clinicians must be prepared to navigate, whether or not formal training has equipped them to do so.

Yet, current educational models fail to bridge the gap between the information age's requirement for knowledge curation and the medical profession's enduring need for deep biomedical competence~\citep{wartman2019empirical, buja2019medical}. While 44\% of physicians report feeling inadequately prepared for new technologies~\citep{minor2020stanford}, educational institutions have struggled to move beyond elective courses and short-term pilots. Existing efforts tend to be fragmented, specialty-specific pilots which show short-term gains but risk knowledge decay. Without longitudinal reinforcement~\citep{vankooten2024framework}, competency frameworks identify what to teach but not how skills should evolve across training~\citep{singla2024developing}, and most training approaches remain anchored to procedural specialties like radiology or ophthalmology~\citep{valikodath2021aao}. The result is a workforce expected to partner with AI tools without the foundational literacy to evaluate, verify, or oversee them.

This leaves a ``Generalist Gap'': a lack of consensus on a comprehensive model for non-procedural or primary care physicians who will increasingly rely on AI-driven decision support. This paper addresses that structural gap. We move beyond the question of \emph{why} AI should be taught to the practical challenge of \emph{how}. To that end, we first clarify several terms that are often conflated in the literature. \emph{Artificial intelligence (AI)} refers broadly to computational systems designed to perform tasks that typically require human cognition. \emph{Machine learning (ML)} is a subset of AI in which models learn patterns from data rather than following explicitly programmed rules. \emph{Generative AI (GenAI)} describes a class of ML models, including large language models (LLMs), capable of producing novel text, images, or other outputs. \emph{Large language models (LLMs)} are a specific GenAI architecture trained on large text corpora to generate and interpret natural language. These distinctions matter for curriculum design: the competencies required to use a diagnostic imaging algorithm differ meaningfully from those needed to evaluate an LLM-generated clinical summary.

By comparing existing models and synthesizing current best practices, we present the \textbf{AI-PACE Framework} (Psychomotor, Affective, Cognitive, Embedded). This model runs longitudinally, introducing AI literacy in the first year of medical school and continuing through practice. Consider that 79\% of U.S. medical schools begin physical exam training within the first two months of the pre-clerkship phase to establish early competency~\citep{uchida2019approaches}. The same logic applies here: early and repeated exposure to AI concepts builds the foundation physicians need to work effectively with these tools as they advance through training. Without this preparation, physicians risk becoming passive consumers of AI outputs rather than professionals equipped to evaluate and apply them appropriately.

\section{Methods}
\label{sec:methods}

To develop a comprehensive framework for AI in medical education, we conducted an integrative review of the literature. Integrative reviews, as defined by Whittemore and Knafl~\citep{whittemore2005integrative}, are specifically designed to synthesize diverse methodological approaches, including empirical and conceptual work, to build new theoretical frameworks. This methodology was selected given our primary objective: not to evaluate the effectiveness of existing curricula, but to map out existing AI competency frameworks and identify structural gaps that a new model could address.

\subsection{Search Strategy}

We searched PubMed, MEDLINE, and ERIC using the following query: (``artificial intelligence,'' ``machine learning,'' ``AI,'' ``data science,'' ``deep learning''), medical education terms (``medical education,'' ``medical student*,'' ``residency,'' ``physician education,'' ``continuing medical education''), and curriculum/competency terms (``curriculum,'' ``curricula,'' ``competenc*,'' ``training program,'' ``educational program'').

The search was limited to publications from 2016 to 2025 to capture the rapid evolution of AI applications in healthcare and the corresponding educational responses. We also queried Asta~\citep{allenai2025asta} to identify papers our initial query might have missed.

Database selection prioritized clinical pedagogy and health professions education literature. Scopus, Web of Science, and IEEE Xplore were not included because our focus was on clinical and educational implementation rather than technical AI development.

\subsection{Study Selection}

The initial search yielded 643 records. We applied six inclusion criteria: (1) focus on AI curriculum or competency frameworks, (2) medical education setting, (3) publication between 2016--2025, (4) English language, (5) peer-reviewed, and (6) focus on US medical programs. The decision to focus on US programs was pragmatic: AI-PACE was developed with the explicit intent of building a pilot course for medical students in the United States, and mapping competencies to the structural phases of US training (UME, GME, CME) was essential for that application. Exclusion criteria included conference abstracts without full text, studies focused on patient care rather than education, and purely technical AI papers without an educational component. Article screening was conducted by a single reviewer (S.P.M.). This is a limitation of the current work and is addressed in the Limitations section. Following full-text review, 23 articles met all inclusion criteria.

\subsection{Data Analysis}

We performed thematic analysis of the 23 papers to identify recurring competencies and curriculum structures. Documents were analyzed for themes across AI competencies, curricular frameworks, educational approaches, and implementation challenges. Competencies were organized across the three domains of Bloom's Taxonomy following Charow et al.~\citep{charow2021artificial}. Curriculum development stages were mapped to established medical education frameworks~\citep{harden1986ten, thomas2022curriculum}.

This analysis (summarized in Table~\ref{tab:frameworks}) revealed specific gaps in existing models, namely, the fragmentation of training into isolated workshops and the under-representation of the affective domain. The AI-PACE Framework operationalizes these competencies by grounding them in the Psychomotor, Affective, and Cognitive domains of Bloom's Taxonomy~\citep{anderson2001taxonomy}, while adding ``Embedded'' as a structural pillar for longitudinal delivery.

\begin{table}[htbp]
  \centering
  \caption{Comparative Analysis of Existing Frameworks}
  \label{tab:frameworks}
  \small
  \begin{tabularx}{\textwidth}{@{}p{2.2cm} X X X X X@{}}
    \toprule
    \textbf{Study / Framework} & \textbf{Audience \& Scope} & \textbf{Methodology} & \textbf{Longitudinal Integration?} & \textbf{Domains Covered} & \textbf{Gaps Identified} \\
    \midrule
    Valikodath et al., 2021~\citep{valikodath2021aao} & Ophthalmology (UME to Fellowship) & Perspective / Task Force Recommendations & Yes (Proposed progression from med school to fellowship) & Fundamentals, surgical simulation, humanization & Specialty-specific; no generalist model \\
    \addlinespace
    van Kooten et al., 2024~\citep{vankooten2024framework} & Radiology (Residency) & Pilot Study / Implementation & No (3-day ``fast-track'' intensive course) & Fundamentals, algorithm building, group ethics & Short-term, specialty-specific; no longitudinal reinforcement \\
    \addlinespace
    Singla et al., 2024~\citep{singla2024developing} & UME (General Medical Students) & Delphi Study (Expert Consensus) & Suggested (Integration into existing curriculum recommended) & Theory, ethics, law, application, communication & Defines what to teach; does not address continuum progression \\
    \addlinespace
    Turner et al., 2025~\citep{turner2025bridging} & PCCM (Pulmonary \& Critical Care) & Perspective / Conceptual & Partial (Mentions ``Scalable Oversight'') & Ethics, robustness, human-AI collaboration & Safe integration principles; not a curriculum design framework \\
    \addlinespace
    Masters, 2020~\citep{masters2020artificial} & General (Admin \& Educators) & Conceptual Framework & Conceptual (Curriculum mapping discussed) & AI literacy, curriculum design & Administrator-focused; limited learner-facing guidance \\
    \addlinespace
    \textbf{AI-PACE Framework} & \textbf{General Medicine (UME, GME, CME)} & \textbf{Integrative Review \& Bloom's Taxonomy} & \textbf{Yes (Spiral curriculum across the full continuum)} & \textbf{P}sychomotor (Workflow); \textbf{A}ffective (Trust); \textbf{C}ognitive (Knowledge); \textbf{E}mbedded (Longitudinal) & \textbf{Addresses the gap}: Generalist, continuum-spanning; explicitly includes Affective domain \\
    \bottomrule
  \end{tabularx}
\end{table}

\section{Current State of AI Education}
\label{sec:current-state}

Despite the proliferation of AI technologies in clinical practice, our analysis reveals a significant lag in the corresponding evolution of medical education frameworks. While recent efforts have begun to define necessary competencies, the field remains fragmented. A scoping review by Tolentino~\citep{tolentino2024curriculum} found only two papers describing comprehensive curriculum frameworks for AI in medicine, with one focused specifically on ophthalmology. A recent scoping review by Rincón~\citep{rincon2025mapping} further confirmed these findings, revealing a lack of standardized AI curriculum frameworks and notable global discrepancies in implementation.

\subsection{Fragmented Implementation and Knowledge Decay}

Current AI education often relies on short-term, intensive ``bootcamps'' or elective workshops. For example, van Kooten~\citep{vankooten2024framework} demonstrated the efficacy of a 3-day AI curriculum for radiology residents, which successfully improved confidence and self-reported knowledge in the short term. However, such concentrated bursts of educational modules risk knowledge decay without reinforcement. They also can fail to integrate AI literacy into the daily clinical workflow where decision-making occurs. The majority of educational interventions consist of isolated courses, workshops, or lectures rather than integrated programs~\citep{charow2021artificial, grunhut2021educating}.

Recent interventions have begun to demonstrate measurable gains. Levingston et al.~\citep{levingston2024theory} developed an AI literacy course for first-year medical students integrating foundational concepts, practical tool use, and ethics. Hopson et al.~\citep{hopson2025enhancing} evaluated a structured 4-week pre-medical AI curriculum using a pre-post design and found significant improvements in AI knowledge scores. These results are encouraging, but both interventions are single-site and lack longitudinal follow-up, reinforcing the need for integrated, continuum-spanning approaches.

\subsection{Specialty Imbalance and the Generalist Gap}

Existing frameworks are heavily skewed toward image-centric specialties. Where AI education does exist, it focuses primarily on these fields~\citep{tolentino2024curriculum}. We discussed previously the efforts focused on ophthalmology, and there have been important principles outlined for Pulmonary and Critical Care Medicine~\citep{turner2025bridging}. Specialty-specific models address depth within defined domains. AI-PACE is designed to address the transferability gap, preparing generalist physicians to evaluate AI-driven decision support across specialties rather than within a single domain.

\subsection{The ``Human'' Gap: Affective Competencies}

Current models also tend to under-address the human element of human-AI interaction. Using a Delphi approach provided a strong consensus on \emph{what} to teach~\citep{singla2024developing}, validating competencies in ethics, theory, and application, but that particular framework focuses primarily on knowledge and skills. What remains under-addressed is the \emph{Affective Domain}: the calibration of trust, the recognition of automation bias, and the preservation of empathy in an AI-mediated patient encounter. Rincón highlighted an emerging educational need for strengthening transversal skills, such as ethical awareness, to support the integration of AI as a practical tool rather than a standalone subject~\citep{rincon2025mapping}.

\section{The AI-PACE Framework}
\label{sec:framework}

\begin{figure}[htbp]
  \centering
  \includegraphics[width=\textwidth]{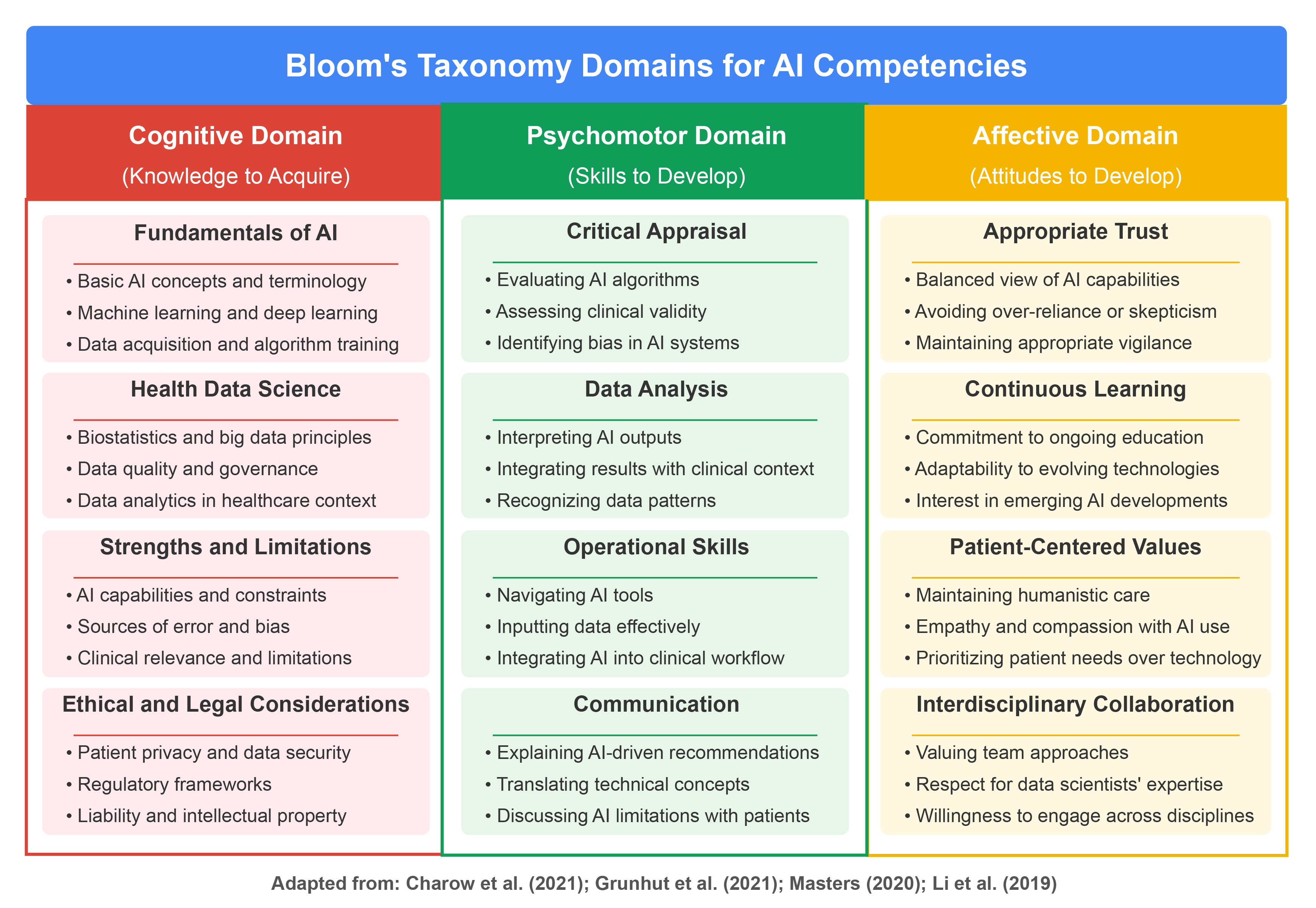}
  \caption{The AI-PACE Taxonomic Framework organizes medical AI competencies across three domains of Bloom's Taxonomy: Cognitive (knowledge), Psychomotor (skills), and Affective (attitudes); with specific competencies mapped to each domain. Adapted from Charow et al.~\citep{charow2021artificial}, Grunhut et al.~\citep{grunhut2021educating}, Masters~\citep{masters2020artificial}, and Li et al.~\citep{li2019fear}.}
  \label{fig:aipace}
\end{figure}

To synthesize our findings and address the structural gaps in current medical education, we developed the \textbf{AI-PACE Framework}. This model organizes competencies into four pillars, grounding the content in the established domains of Bloom's Taxonomy (Psychomotor, Affective, Cognitive) while introducing a fourth structural dimension (Embedded) to ensure longitudinal integration across the medical training continuum (Figure~\ref{fig:aipace}).

We start by examining the mapping to Bloom's taxonomy and then move on to the embedded approach.

\subsection{Psychomotor Domain (Skills \& Workflow)}

The Psychomotor domain focuses on the shift from theoretical knowledge to clinical utility, emphasizing the ``hands-on'' ability to navigate tools, interpret probabilistic outputs, and communicate these findings to patients.

\begin{itemize}
  \item \textbf{Operational Skills:} Navigating AI tools in clinical workflows, inputting data effectively, and retrieving meaningful results~\citep{tolentino2024curriculum}.
  \item \textbf{Critical Appraisal:} Evaluating AI algorithms for clinical validity, bias, and applicability to specific patient populations~\citep{charow2021artificial}. This includes interpreting key performance metrics (sensitivity, specificity, AUC) without necessarily calculating them.
  \item \textbf{Data Analysis:} Interpreting outputs from AI systems and integrating them into clinical decision-making~\citep{charow2021artificial}.
  \item \textbf{Communication:} Explaining AI-driven recommendations to patients in understandable terms and translating technical concepts~\citep{masters2020artificial}.
\end{itemize}

\subsection{Affective Domain (Appropriate Entrustment \& Trust Calibration)}

The affective domain encompasses the attitudes, values, and emotional intelligence required for human-AI partnership. This domain is central to addressing the ``human gap'' identified in our analysis. However, framing this as ``trust'' undersells what clinical training actually requires. Clinicians must learn to assess AI outputs as they would a junior colleague's work. This means understanding where the tools can fail (i.e., LLMs summarize effectively but struggle to extract qualitative meaning from patient narratives). Therefore, curricula must address these boundaries explicitly rather than treating AI competence as a binary trust decision.

Trust calibration in this context carries a duality that curricula must acknowledge. Clinicians must learn to evaluate the trustworthiness of AI outputs (trust in the AI), but educators must simultaneously assess whether the trainee has developed sufficient clinical judgment to make that evaluation reliably (trust in a trainee's competencies). Gin et al.\ propose applying the Entrustable Professional Activity (EPA) framework to AI tools as trustees, assessing AI across ability, integrity, and benevolence in parallel with how we assess trainees, as a structured approach to operationalizing both sides of this trust relationship~\citep{gin2025entrustment}.

\begin{itemize}
  \item \textbf{Appropriate Trust Calibration:} Developing a balanced perspective on AI capabilities to avoid both ``automation bias'' (over-reliance) and ``algorithm aversion'' (unwarranted skepticism)~\citep{grunhut2022needs}.
  \item \textbf{Patient-Centered Values:} Maintaining focus on humanistic care while leveraging technological tools~\citep{li2019fear}. AI should augment rather than replace empathy and compassion.
  \item \textbf{Collaboration:} Valuing interdisciplinary teamwork with data scientists and engineers~\citep{charow2021artificial}.
  \item \textbf{Continuous Learning:} Fostering commitment to ongoing education as AI technologies evolve~\citep{wartman2018medical}.
\end{itemize}

\subsection{Cognitive Domain (Knowledge)}

The cognitive domain establishes the foundational knowledge base required to understand AI as a tool.

\begin{itemize}
  \item \textbf{Fundamentals of AI:} Understanding basic concepts, terminology (machine learning, deep learning), and knowing that ML models generate probabilistic predictions based on patterns, not rigid rules~\citep{masters2020artificial, paranjape2019introducing}. Foundational training must also update outdated skepticism. The 2022-era critique of AI as ``stochastic parrots''---probabilistic word repeaters---no longer reflects current model capabilities~\citep{qiu2025quantifying, mccoy2025assessment, anthropic2025claude45, deepmind2025gemini3}. Reasoning and agentic functions have advanced substantially. Curricula should ground skepticism in actual technical limitations.
  \item \textbf{Health Data Science:} Comprehending biostatistics, big data principles, and data governance~\citep{charow2021artificial}.
  \item \textbf{Strengths and Limitations:} Recognizing the capabilities and constraints of AI systems, including sources of error and clinical relevance~\citep{grunhut2021educating}.
  \item \textbf{Ethical and Legal Considerations:} Understanding patient privacy, algorithmic bias, and regulatory frameworks governing AI in healthcare~\citep{masters2020artificial, pregowska2024artificial}.
\end{itemize}

\subsection{Embedded (Longitudinal Integration)}

The ``Embedded'' pillar addresses the structural limitations of fragmented bootcamp models by distributing AI competencies across the full medical training continuum, which aligns with the higher levels of Harden's ``integration ladder''~\citep{harden2000integration}. Rather than treating AI as an isolated elective or a standalone ``AI week,'' this approach aligns with Han's work~\citep{han2019medical} by weaving AI literacy into the existing fabric of pre-clerkship and clerkship coursework. This longitudinal model directly bridges the ``Generalist Gap'' by establishing foundational cognitive and affective competencies early in training, which then branch into specialized psychomotor skills during residency and independent practice.

Implementation must also address the faculty gap. Early-career faculty often have strong interest in AI literacy but may hesitate to enter classes where students could outpace them. This creates a disconnect: the faculty best positioned to integrate emerging tools might also be those most wary of the optics of learning along with their students. Therefore, parallel upskilling pathways for junior faculty are essential; without them, we cannot expect faculty to model the critical evaluation and entrustment behaviors the curriculum intends to teach.

\begin{itemize}
  \item \textbf{Undergraduate Medical Education (UME):} Focuses on foundational concepts and basic ethical awareness. AI-specific content is integrated into existing modules; for example, the cognitive distinction between deterministic algorithms and probabilistic machine learning is taught alongside Evidence-Based Medicine (EBM) and Biostatistics.
  \item \textbf{Graduate Medical Education (GME):} Shifts toward specialty-specific applications and workflow integration. Residents move from theory to clinical utility, practicing ``human-in-the-loop'' verification during specialty rotations (e.g., Internal Medicine or Radiology clerkships) through hands-on, case-based learning.
  \item \textbf{Continuing Medical Education (CME):} Emphasizes practice integration strategies, emerging technologies, and institutional leadership. Targeted workshops and self-directed learning address the evolving needs of practicing physicians, focusing on the selection, implementation, and oversight of AI tools within active workflows.
\end{itemize}

To operationalize this approach, Table~\ref{tab:curriculum-map} provides a curriculum map that aligns the AI-PACE domains with specific milestones and existing integration points across the medical education continuum. This map transitions from foundational theory in undergraduate years to advanced clinical oversight and leadership in continuing practice. It outlines the Embedded (Longitudinal) pillar of AI-PACE by mapping the Cognitive, Psychomotor, and Affective domains to specific milestones in training.

\begin{table}[htbp]
  \centering
  \caption{AI-PACE Curriculum Map and Integration Strategy}
  \label{tab:curriculum-map}
  \small
  \begin{tabularx}{\textwidth}{@{}p{2.6cm} p{1.9cm} X X X@{}}
    \toprule
    \textbf{Training Phase} & \textbf{Primary Focus} & \textbf{Cognitive (Knowledge)} & \textbf{Psychomotor (Skills/Workflow)} & \textbf{Affective (Attitudes/Trust); Integration Point \& Learning Approach} \\
    \midrule
    Undergraduate Foundation (UME --- Pre-Clinical) & Foundations & AI terminology \& concepts; ML vs.\ deterministic logic; probabilistic nature of AI & Basic prompt engineering for study aids; navigating literature search & Recognizing automation bias; valuing data privacy; skepticism of ``black boxes.'' \newline \textbf{Introductory courses \& integrated modules} in Epidemiology, Biostatistics, \& Health Systems Science. \\
    \addlinespace
    Undergraduate (UME --- Clinical) & Exposure & Interpreting AI metrics (AUC, sensitivity); regulatory landscapes (FDA); data science basics & Critical appraisal of AI tools during rounds; using AI for differential diagnosis; communicating results to patients & Calibration of trust; empathy in AI-mediated patient encounters. \newline \textbf{Clinical clerkships} (e.g., IM, Radiology) via integrated modules and clinical exposure. \\
    \addlinespace
    Graduate (GME --- Residency) & Application & Specialty-specific AI applications; sepsis predictors; image triage algorithms & Clinical integration \& decision making; ``human-in-the-loop'' verification; EHR documentation; error reporting & Professional accountability; collaboration with data science teams. \newline \textbf{Hands-on applications \& case-based learning} during morning reports, grand rounds, \& QI projects. \\
    \addlinespace
    Continuing (CME --- Practice) & Mastery \& Leadership & Emerging LLM capabilities; quality \& performance assessment; liability updates & Workflow optimization; selecting/implementing AI tools; leadership in AI implementation & Fostering a culture of safety; mentoring junior staff on AI oversight. \newline \textbf{Targeted workshops, just-in-time training}, and institutional leadership roles. \\
    \bottomrule
  \end{tabularx}
\end{table}

\subsection{Integrated Pedagogical Methods}

\begin{itemize}
  \item \textbf{Case-Based Learning:} Use specialty-specific ``AI-Fail'' case studies to teach trust calibration and verification.
  \item \textbf{Simulation Objective Structured Clinical Examinations (OSCEs):} Practice explaining AI-generated diagnoses and algorithmic limitations to standardized patients.
  \item \textbf{Quality Improvement Projects:} Encourage GME/CME learners to evaluate AI tool performance and safety within their actual clinical workflows.
\end{itemize}

\section{Implementation Strategy}
\label{sec:implementation}

\begin{figure}[htbp]
  \centering
  \includegraphics[width=\textwidth]{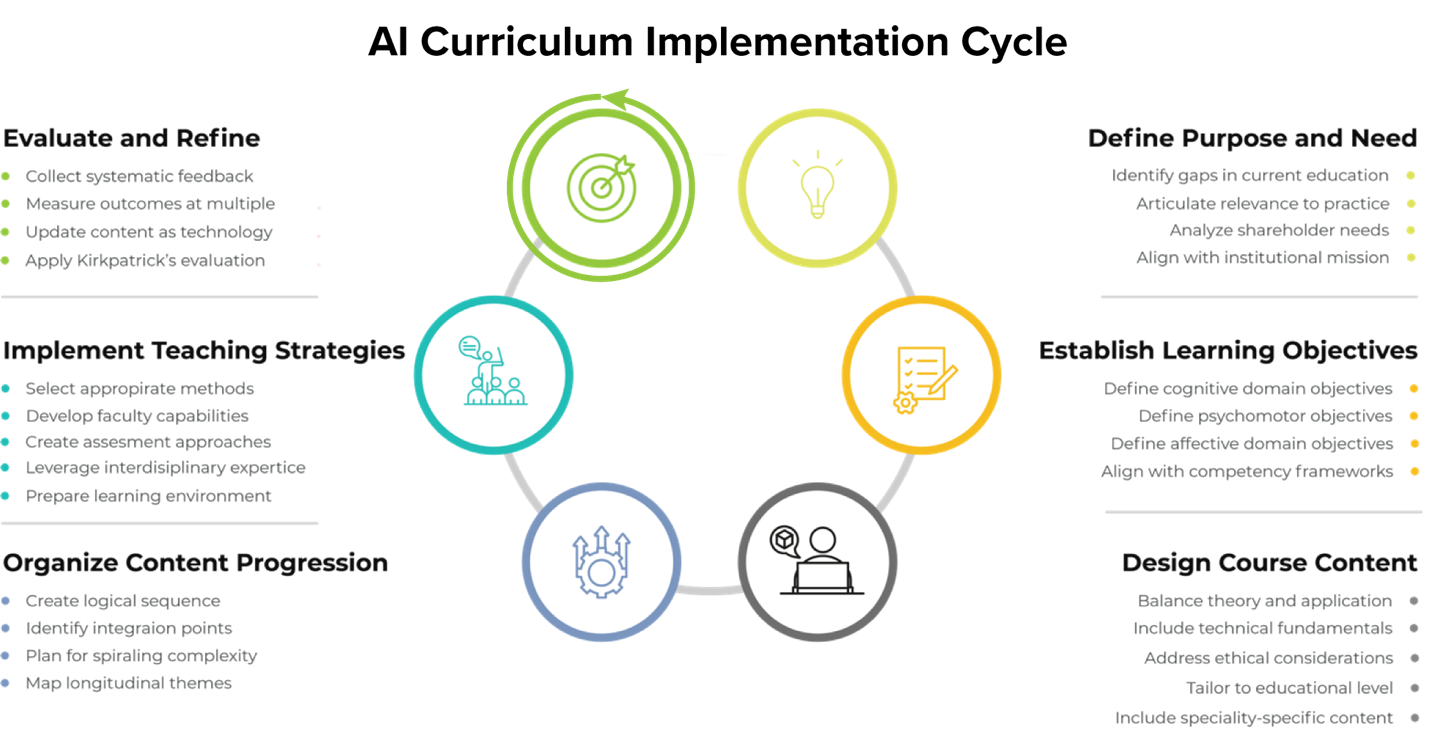}
  \caption{A six-step implementation cycle for AI curriculum development. Steps progress from defining purpose, establishing objectives, designing/organizing content, teaching strategies, and evaluation. The evaluation step is continuous rather than terminal, reflecting the accelerated pace at which AI capabilities and clinical applications evolve.}
  \label{fig:implementation}
\end{figure}

Successfully deploying the AI-PACE framework requires a structured approach to curriculum development that can keep pace with rapid technological shifts. Based on established models by Harden~\citep{harden1986ten} and Thomas~\citep{thomas2022curriculum}, we put forward a six-step cycle specifically adapted for the dynamic nature of clinical AI (Figure~\ref{fig:implementation}).

\begin{enumerate}
  \item \textbf{Define Purpose and Need:} Identify gaps in current education and articulate relevance to future practice.
  \item \textbf{Establish Learning Objectives:} Define specific outcomes across the Cognitive, Psychomotor, and Affective domains (as outlined in AI-PACE).
  \item \textbf{Design Course Content:} Balance theoretical foundations with practical applications, tailoring content to the educational level (UME vs GME).
  \item \textbf{Organize Content Progression:} Create a logical sequence building from foundational to advanced concepts, identifying integration points within the existing curriculum.
  \item \textbf{Implement Teaching Strategies:} Select appropriate methods (lectures, case-based learning, projects) and develop faculty capabilities.
  \item \textbf{Evaluate and Refine:} Collect systematic feedback and measure outcomes. Crucially, this step must be continuous to update content as technology evolves.
\end{enumerate}

\section{Discussion}
\label{sec:discussion}

The rapid evolution of AI in medicine has outpaced the development of comprehensive educational frameworks, creating a ``preparedness gap'' that threatens the safe integration of these technologies into clinical practice. The AI-PACE framework is designed to address these limitations through a spiral curriculum model that evolves with the learner.

\subsection{The ``Integration vs.\ Addition'' Debate}

A primary barrier to AI education is the ``crowded curriculum'' problem. Medical school and residency schedules are already saturated~\citep{wartman2019empirical}. Integrating AI concepts into existing courses rather than creating new courses can help address this challenge~\citep{grunhut2022needs}. For example, the \emph{cognitive} distinction between deterministic algorithms and probabilistic machine learning should be taught alongside EBM during pre-clerkship years. Similarly, the \emph{psychomotor} skill of ``human-in-the-loop'' verification should be practiced during existing clinical clerkships, explicitly avoiding the ``fragmented bootcamp'' model that detaches technology from the daily clinical workflow. This integration model mirrors how EBM itself was once a novel addition that is now woven into the fabric of clinical reasoning.

At the University of California, Davis School of Medicine, Health System Science was introduced as a new longitudinal pillar through the I-Explore curriculum reform folded into existing courses rather than added as standalone credit hours~\citep{ucdavis2021iexplore}. AI literacy could follow the same path, with cognitive foundations embedded in Biostatistics and EBM coursework, psychomotor skills practiced during existing clinical clerkships, and affective competencies addressed through the documentation and supervision structures already present in clinical training. At the GME level, Preiksaitis demonstrates that supervising resident AI use can be embedded into existing case presentations and clinical rounds through structured prompts, without requiring dedicated curriculum time~\citep{preiksaitis2026supervising}.

\subsection{The Resource Challenge: Faculty and Funding}

Implementing AI-PACE requires three categories of institutional investment. First, faculty development: many medical educators lack expertise in AI and data science~\citep{grunhut2022needs, chan2019applications}, which can be addressed through structured upskilling pathways, recruitment of clinician-informaticians, and partnerships with engineering curricula~\citep{briganti2020artificial}. Second, interdisciplinary infrastructure: breaking down silos between Computer Science, Informatics, and Medicine is necessary to keep content current as AI capabilities evolve rapidly~\citep{charow2021artificial, masters2020artificial}; institutions with existing health informatics graduate programs may already have this partially in place. Third, technical access: students frequently lack the same licensed access to AI tools available to clinical staff, as licensing costs and IT compliance support are typically budgeted for employees rather than trainees. Proactive coordination between medical school administration and health system IT governance is required to close that gap.

\subsection{Validation and Future Directions}

As medical education adopts AI curricula, we must rigorously validate their impact. Future research should move beyond self-reported confidence (Kirkpatrick Level 1) to measure changes in behavior (Level 3) and patient outcomes (Level 4)~\citep{kirkpatrick2006evaluating, yardley2012kirkpatrick}. Does an AI-literate resident make safer decisions when using a diagnostic support tool? By standardizing the educational model through AI-PACE, we can begin to generate the multi-institutional data needed to answer questions like this. This work is also timely given emerging evidence that relying on GenAI for efficiency may come at a cost of skill formation~\citep{shen2026ai}. This is particularly relevant given emerging concerns about deskilling, mis-skilling, and never-skilling as physicians increasingly offload cognitive tasks to AI tools~\citep{natali2025deskilling, berzin2025preserving}.

To move from framework to evidence, we are currently piloting AI-PACE through a 4-week elective course for medical students at the University of California, Davis. Pre- and post-intervention AI preparedness will be measured using the Medical Artificial Intelligence Readiness Scale for Medical Students (MAIRS-MS), a validated instrument developed for this purpose~\citep{karaca2021mairs}. Results from this pilot are intended as a follow-up case study to this paper. Beyond the US context, international adaptation of AI-PACE represents an important next step, one that would likely require a broader consensus methodology, such as a Delphi study, to account for the variation in training structures and accreditation requirements across health systems.

Neither the Liaison Committee on Medical Education (LCME) nor the Accreditation Council for Graduate Medical Education (ACGME) currently have binding AI-specific standards, but the field is moving in that direction. A 2025 Josiah Macy Jr.\ Foundation conference co-sponsored by AAMC and ACGME explicitly recommended implementing AI curricula and competencies and modifying accreditation processes to account for AI~\citep{macy2025conference}. AI-PACE is designed to map onto this trajectory, providing institutions with a structured framework for responding to emerging accreditation expectations rather than waiting for prescriptive requirements that do not yet exist.

\subsection{Limitations}

Several limitations of this work should be acknowledged. First, article screening was conducted by a single reviewer (S.P.M.), which introduces the potential for selection bias. A formal systematic review with blinded, multi-reviewer screening was outside the scope of this framework development effort. Second, the review was intentionally restricted to medical training programs in the United States, reflecting the immediate practical need that motivated this work: developing a pilot course for US medical students. This US-centric focus limits the generalizability of AI-PACE to international contexts, where training structures, accreditation requirements, and pedagogical approaches can vary considerably. There is the potential for international adaptation of the framework, which could be achieved through a consensus methodology such as a Delphi study. Third, this manuscript cites formal technical reports and model cards from frontier AI laboratories alongside peer-reviewed literature when characterizing current model capabilities. This reflects a structural limitation of the field: the 9--12-month peer-review publication cycle means that empirical evaluations of LLM performance often assess models one or two generations behind those currently in clinical use. System cards and model cards, while not peer-reviewed, represent the timeliest available documentation of capability boundaries and safety evaluations for recently released models. We acknowledge this as a methodological constraint and encourage readers to weigh these sources accordingly.

\section{Conclusion}
\label{sec:conclusion}

The integration of AI into healthcare necessitates corresponding changes in medical education. Our analysis of the literature reveals a significant gap between the rapid adoption of AI in clinical practice and the preparation of physicians to use these technologies effectively. While a limited number of educational initiatives exist, comprehensive curriculum frameworks are largely absent or fragmented. We have identified competencies across cognitive, psychomotor, and affective domains that should guide AI education for medical professionals. Effective curricula should balance technical knowledge with clinical applications and ethical considerations, be integrated longitudinally throughout medical training, and involve interdisciplinary collaboration.

The AI-PACE framework presented in this paper offers a starting point for medical educators seeking to develop AI curricula. Future research should focus on implementing and evaluating such curricula, developing specialty-specific educational approaches, and creating faculty development programs to support AI education. As AI continues to transform healthcare, ensuring that physicians are adequately prepared to leverage these technologies while maintaining humanistic, patient-centered care is among the most pressing challenges facing medical education.

\section*{Acknowledgements}

This work was supported by a grant provided to CITRIS Health at UC Berkeley from The First Affiliated Hospital, Sun Yat-Sen University (grant number A25-2059-002), and by the National Center for Advancing Translational Sciences, National Institutes of Health (grant number UL1 TR001860).

S.P.M.\ thanks Melanie L.\ McGrath, PhD, LAT, ATC, for her assistance with developing the manuscript in its earliest form and for her steadfast support.

\section*{Author Contributions}

S.P.M.\ conceptualized the framework, wrote the original draft, and designed the figures. K.K.K., K.J., H.W., and N.A.\ critically reviewed and edited the manuscript. All authors have read and approved the final manuscript.

\section*{Competing Interests}

The authors declare no competing interests.

\bibliographystyle{unsrtnat}
\bibliography{references}

\end{document}